\documentclass{article}
\usepackage[symbol]{footmisc}
\usepackage{amsmath}
\usepackage{amsfonts}
\usepackage{amssymb}
\usepackage{hyperref}
\newcommand{\onlinecite}[1]{\cite{#1}}
\newcommand{\mycite}[1]{~\cite{#1}}
\newcommand{\myonlinecite}[1]{Ref.\onlinecite{#1}}
\newcommand{\runninghead}[2]{}
\newcommand{\address}[1]{}
\newcommand{\D}{\,\mathrm{d}}
\newcommand{\pd}[2]{\ensuremath{\frac{\partial #1}{\partial #2}}}
\newcommand{\pdl}[2]{\ensuremath{\partial #1 / \partial #2}}

\newcommand{\ie}{\textit{i.e.},~}
\newcommand{\eg}{\textit{e.g.},~}
\newcommand{\etc}{\textit{etc.}}
\newcommand{\vs}{\ensuremath{v_\text{s}}}
\newcommand{\rs}{\ensuremath{\rho_\text{s}}}
\newcommand{\rn}{\ensuremath{\rho_\text{n}}}
\newcommand{\vn}{\ensuremath{v_\text{n}}}
\newcommand{\vb}{\ensuremath{\mathbf{v}}}
\newcommand{\wb}{\ensuremath{\mathbf{w}}}
\newcommand{\vbd}{\ensuremath{\mathbf{\dot{v}}}}
\newcommand{\wbd}{\ensuremath{\mathbf{\dot{w}}}}
\newcommand{\vsb}{\ensuremath{\mathbf{v}_\text{s}}}
\newcommand{\vnb}{\ensuremath{\mathbf{v}_\text{n}}}
\newcommand{\vsbd}{\ensuremath{\mathbf{\dot{v}}_\text{\rm s}}}
\newcommand{\vnbd}{\ensuremath{\mathbf{\dot{v}}_\text{\rm n}}}
\newcommand{\avrg}[1]{\ensuremath{\left<#1\right>}}

\author{L.A.\,Melnikovsky\footnote{E-mail: leva@kapitza.ras.ru}}
\address{P.L.\,Kapitza Institute for Physical Problems, Moscow, Russia}
\title{Polarization of Dielectrics by Acceleration}
\runninghead{L.A.\,Melnikovsky}{Polarization of Dielectrics by Acceleration}
\begin{document}
\maketitle
\begin{abstract}
We argue that acceleration induces electric polarization in usual
dielectrics. Both accelerations in superfluid (\vsbd\ and
\vnbd ) participate in the medium polarization. Excitations
contribution to the polarization is calculated at low temperatures.
Estimates of the effect show order of magnitude agreement with recent
experimental results on electric effect of superflow.
\end{abstract}

\section{Introduction}
Excitation of electric currents by metal acceleration is well known and
usually associated with the Stewart-Tolman\mycite{TS} effect. Similar
phenomenon exists in dielectrics. In this paper we show that
acceleration polarizes dielectrics. There are two formally separated
mechanisms of the effect. Individual accelerated molecules acquire
intrinsic dipole moment. This contribution is discussed in
Sec.\ref{gravel} Collective polarization caused by the medium
inhomogeneity required for the accelerated motion is considered in
Sec.\ref{flexel}

As reported\mycite{ryba} by A.S.\,Rybalko, a second-sound wave gives rise
to an electric field in superfluid.\footnote{Subsequent
work\mycite{torsion} was aimed to discover electric effect of the
superfluid flow in torsional oscillator. There has also been an attempt
to describe these experimental results
phenomenologically\mycite{kosevich}.} In Sec.\ref{super} we show that
electric effect of superflow can be interpreted as a manifestation of
the acceleration-induced polarization. Eventually in Sec.\ref{discuss}
we estimate the effect magnitude in the second-sound wave.

\section{Gravitoelectric effect}
\label{gravel}
Applied uniform electric field $\mathbf{E}$ polarizes isotropic dielectric, in
linear approximation its polarization (dipole moment per unit volume) is
given\mycite{LL8} by the equation
\begin{equation}
\label{PofE}
\mathbf{P}=n\mathbf{d}=\varkappa \mathbf{E},
\end{equation}
where $\mathbf{d}$ is a single atom dipole moment, $n$ is the atom
density, $\varkappa=(\varepsilon -1)/4\pi$ is the electric susceptibility,
and $\varepsilon$ is the permittivity of the medium. In presence of an
external gravity field $\mathbf{g}$ equation \eqref{PofE} should be
extended to include additional term
\begin{equation}
\label{PofEG}
n\mathbf{d}=
\varkappa \mathbf{E} + \mathbf{P}_g=
\varkappa \mathbf{E} + \gamma \mathbf{g}.
\end{equation}

Phenomenologically we can not calculate the ``gravitoelectric
susceptibility'' $\gamma$, it is an intrinsic property of the
dielectric. We can, however, estimate it from microscopic
analysis. Consider the dipole moment $\mathbf{d}$ of a
single atom. Both electric and gravity forces acting upon it urge to
tear electric cloud away from the nucleus. The polarization must
therefore be equal to zero if these two effects compensate each other.
Electric ``tear force'' (the difference between the forces acting on the
nucleus and on the electrons) is approximately $F_e\sim 2EZe$, where $Z$
is the atom number and $e$ is the elementary charge. Gravity effectively
acts on the nucleus only, appropriate ``force'' is $F_g\sim gM$, where
$M$ is the atom mass. Equating these two ``forces'' we obtain
$\gamma \sim \varkappa M /(2Ze)$.

So far we considered the dielectric at rest. Influence of the velocity,
unless it is comparable to the speed of light, can be neglected, but
equivalence principle of the General Relativity dictates how the
acceleration \vbd\ enters \eqref{PofEG}
\begin{equation}
\label{PofEGA}
\mathbf{P}_g=\gamma (\mathbf{g}-\vbd).
\end{equation}

\section{Flexoelectric effect}
\label{flexel}
Uniform medium polarization is a single atom dipole moment times the
atom density \eqref{PofE}. In a nonuniform medium
(and even uniform gravity field breaks the system homogeneity)
higher order electric
moments (quadrupole \etc) also contribute to the macroscopic
polarization $\mathbf{P}=n\mathbf{d}+\mathbf{P}_f$. These contributions
should be separable since they depend differently on external field and
its gradients\mycite{frenkel}. Similar phenomenon is known, \eg in
liquid crystals and is referred to as quadrupole {\em flexoelectric}
effect\footnote{Possible manifestation of the effect in
superfluid ${}^3$He vortices is discussed in~\myonlinecite{volovik}.}\mycite{flex}.

To find the quadrupole contribution to the polarization we follow the
arguments used in \myonlinecite{LL8} to introduce the polarization itself. If
the medium has no external charges nor the atoms posses any
intrinsic dipole moment, then the total dipole moment of a body of any
shape is identically zero
\begin{equation*}
\int\mathbf{P}_f \D V=0.
\end{equation*}
Consequently, the polarization can be expressed as a divergence of some
tensor
\begin{equation*}
P^i_f =-\frac{1}{6}\pd{Q^{ij}}{x^j}
\end{equation*}
and $Q^{ij}$ is zero outside the body. To elucidate the meaning of
$Q^{ij}$ consider the total ``generalized\footnote{Conventional
quadrupole moment tensor is traceless.} quadrupole moment'' of the body
$\mathfrak{Q}^{ij}$
\begin{equation*}
\mathfrak{Q}^{ij}
 = 3\int x^i x^j \varrho \D V
 = -3\int x^i x^j \pd{P^k_f}{x^k} \D V
% = 3\int (\delta^{ki} x^j + \delta^{kj} x^i) P^k \D V
% = -\frac{1}{2} \int (\delta^{ki} x^j + \delta^{kj} x^i) \pd{Q^{lk}}{x^l} \D V
% = \frac{1}{2} \int (\delta^{ki}\delta^{lj} + \delta^{kj}\delta^{li}) Q^{lk} \D V
 = \int Q^{ij} \D V.
\end{equation*}
Here $\varrho=-\nabla \mathbf{P}_f$ is the charge density and $Q^{ik}$ is assumed
to be symmetric without loss of generality. Tensor $Q^{ij}$ is therefore
a ``generalized quadrupole moment'' density per unit volume.

Macroscopic electric field $\mathbf{E}=\mathbf{D}-4\pi\mathbf{P}$ is the
average microscopic field $\mathbf{E}=\avrg{\mathbf{e}}$ and can be
detected within the medium by a probe charge. To find the
field outside the medium, the Maxwell equations should be equipped with
the boundary conditions.
They should be corrected if bulk quadrupole moment is taken into account.
Appropriate contribution to the contact potential
for the interface between the medium 
and vacuum (denoted below by $m$ and $v$ respectively)
is determined by the double layer at the medium surface.
To find it consider the integral along an
infinitesimal path $m \rightarrow v$ crossing the surface:
\begin{equation}
\label{boundary}
\varphi_v - \varphi_m = -\int\limits_m^v \mathbf{E} \D \mathbf{s}
=  4\pi \int\limits_m^v P^i_f \D x^i
= -\frac{2\pi}{3} \int\limits_m^v \pd{Q^{ij}}{x^j} \D x^i
= \frac{2\pi}{3} Q^{ij} n^j n^i.
\end{equation}
Here $\mathbf{n}$ is a unit vector normal to the surface and
directed outwards the medium. 

For an isotropic medium (normal liquid) $Q^{ik}=Q\delta^{ik}$ and can be
estimated as $Q\sim -Ze r_0^2\rho/M$,
where $r_0$ is the atom radius and $\rho$ is the liquid density.
This gives for the polarization
\begin{equation}
\label{PofRHO}
\mathbf{P}_f
\sim \frac{Ze}{6} \nabla \left(\frac{M^{2/3}}{4\rho^{2/3}}\frac{\rho}{M}\right)
%= \frac{Ze}{24M^{1/3}}\nabla \left(\rho^{1/3}\right).
\sim \phi \nabla \left(\rho^{1/3}\right),
%=-\frac{Ze}{72M^{1/3}\rho^{2/3}}\nabla \rho
%=-\frac{Ze\rho^{1/3}}{72M^{1/3}c^2} \mathbf{g}.
\end{equation}
where $\phi = Ze \left/ \left(24M^{1/3}\right) \right. $.

\section{Polarization of Superfluid}
\label{super}
Superfluid may posses two types of motion: superfluid and normal flows
characterized by the velocities \vsb\ and \vnb, and  by the densities
\rs\ and \rn. Equations \eqref{PofEGA} and
(\ref{boundary},\ref{PofRHO}) therefore require generalization to this
two-fluid case.

Phonons, the most important excitations at low temperatures, can be
described by macroscopic hydrodynamics.\mycite{LL6}\footnote{Similar
approach is used in \myonlinecite{birefringence} to calculate optical
anisotropy in superfluid.}
Equation
\eqref{PofEGA} when averaged over thermal phonons gives macroscopic
gravitoelectric polarization as $\mathbf{P}_g=-\avrg{\gamma\vbd}$.
Microscopic acceleration is obtained from the Euler equation
\begin{equation}
\label{da}
\vbd \equiv \frac{\D\vb}{\D t} \equiv
  \frac{\partial \vb}{\partial t} 
+ v^i \frac{\partial \vb}{\partial x^i}=
-\frac{\nabla \mathfrak{p}
}{\rho},
\end{equation}
where $\mathfrak{p}(\rho)$ is the microscopic pressure. 
%fixme - rho is already defined??? 
%and $\rho$ is the density of the liquid.
Electric susceptibility of helium is very small and is therefore
proportional to the density. This also applies to the gravitoelectric
susceptibility $\gamma \propto \varkappa \propto \rho$. From \eqref{da}
we get
\begin{equation}
\label{macroP1}
\mathbf{P}_g
%=-\avrg{\gamma \vbd}
=\frac{\gamma}{\rho} \avrg{\nabla \mathfrak{p}}=
\frac{\gamma}{\rho}\pd{}{x^i}\avrg{\Pi^{ik}-\rho v^i v^k}=
\frac{\gamma}{\rho}\pd{}{x^i}\left(\Pi^{ik}-\avrg{\rho v^i v^k}\right),
\end{equation}
where $\Pi^{ik}$ is the momentum flux density tensor. Its microscopic
and average macroscopic values are given by
\begin{equation}
\label{PI}
\Pi^{ik}=\mathfrak{p}\delta^{ik}+\rho v^iv^k
\quad\text{and}\quad
\Pi^{ik} = \vs^i
j^k+\vn^k j^i-\rho \vn^k \vs^i+p\delta_{ik}
\end{equation}
respectively, where $\mathbf{j}$ is the mass flux\footnote{We use  the
same symbols for the microscopic and the average values of such
quantities as density $\rho$ and momentum flux $\Pi^{ik}$, while keep
$\mathfrak{p}$ and $p$ distinguishable since $\avrg{\mathfrak{p}}\ne
p$.} and $p$ is the pressure. Anisotropic part of \avrg{\rho v^i v^k} is
quadratic with respect to the relative velocity $\wb=\vnb-\vsb$, in linear
approximation it can be neglected $\avrg{\rho v^i
v^k}=\delta^{ik}\avrg{\rho v^2}/3$. In phonons mean kinetic energy
%\begin{equation*}
$E_k=\avrg{\rho v^2/2}$
%\end{equation*}
is equal to mean potential energy
\begin{equation}
\label{EP}
E_p=\frac{1}{2\rho}\pd{p}{\rho}\avrg{(\rho-\avrg{\rho})^2}=
\frac{c^2}{2\rho}\avrg{(\rho-\avrg{\rho})^2},
\end{equation}
where $c$ is the velocity of sound\footnote{No distinction is made between
isothermal and adiabatic compressibility $\pdl{p}{\rho}=c^2$ nor between
heat capacity $C$ at constant pressure and density.}. In other words
$\avrg{\rho v^2}=\delta E_0$, where $\delta$ designates the
increment
with respect to the value
at $T=0$ (neglecting the ground state energy),
\ie $\delta E_0 = E_0-\left. E_0\right|_{T=0}$. Here
$E_0=T\sigma\rho+\mu\rho-p$ is the energy density in the frame of
reference of the superfluid component, $\sigma$ and $\mu$ are the
entropy and the chemical potential per unit mass.
%Neglecting higherorder terms
%\begin{multline}
\begin{equation}
\label{DE}
\D E_0 = T\D (\rho\sigma) + \mu\D\rho=
T\rho\D\sigma + \frac{p+E_0}{\rho}\D\rho=
%C\rho\D T + W\pd{\rho}{p}\D p=
C\rho\D T + \frac{W}{c^2}\D p,
\end{equation}
%\end{multline}
where $W=(p+E_0)/\rho$ is the heat function per unit mass. Finally
combining \eqref{macroP1}, \eqref{PI}, and \eqref{DE} we get
\begin{equation}
\label{PofPT}
\mathbf{P}_g=
\frac{\gamma}{\rho}
\left(1 - \frac{\delta W}{3c^2}\right)\nabla p-
\frac{\gamma C}{3}\nabla T=
\frac{\gamma}{\rho}
\left(1 - \frac{\delta W}{3c^2} - \frac{C}{3\sigma}\right)\nabla p+
\frac{\gamma C}{3\sigma}\nabla \mu.
\end{equation}
This can also be expressed with the help of linearized superfluid
hydrodynamics equations as
\begin{equation}
\mathbf{P}_g=
\frac{\gamma C\rn}{3\rho\sigma}\wbd
-\frac{\gamma}{\rho}
\left(1 - \frac{\delta W}{3c^2}\right)\left(\rs\vsbd + \rn\vnbd\right)=
-\gamma_\text{n}\vnbd
-\gamma_\text{s}\vsbd,
\end{equation}
where
\begin{equation}
\gamma_\text{n}=\frac{\gamma\rn}{\rho}\left(1-\frac{\delta W}{3c^2} -
                                      \frac{C}{3\sigma} \right),
\quad\quad
\gamma_\text{s}=\frac{\gamma}{\rho}\left(\rs - \frac{\rs\delta W}{3c^2} +
                               \frac{ C\rn}{3\sigma}  \right).
\end{equation}

Similarly, averaged \eqref{PofRHO} reads
\begin{equation*}
\mathbf{P}_f
\sim \phi \nabla \avrg{\rho^{1/3}}
=\phi \nabla \left(\rho^{1/3} 
            - \frac{1}{9\rho^{5/3}} \avrg{(\rho-\avrg{\rho})^2}
	    \right).
\end{equation*}
Employing \eqref{EP} we get for the phonons
\begin{equation}
\label{PofRHON}
\mathbf{P}_f
\sim
\phi \nabla \left(\rho^{1/3} 
            - \frac{1}{9c^2\rho^{2/3}} \delta E_0
	    \right)
=
\phi \nabla \left(\rho^{1/3} 
            - \frac{\rn}{12\rho^{2/3}}
	    \right),
\end{equation}
using $\rn=4\delta E_0\left/\left(3c^2\right)\right.$ for the phonon
term in the normal density\mycite{LL9}.

Rotons, unlike phonons, being the short-wavelength excitations, are not
described by macroscopic hydrodynamics. Their contribution to the
quadrupole moment is not necessarily scalar,
\ie ${\cal Q}^{ij}=Q^{ij}-\left. Q^{kk}\delta^{ij} \right/ 3\not\equiv 0$.
Contribution of a single roton to ${\cal Q}^{ij}$
can be estimated (see \eqref{PofRHON})
\begin{equation*}
q^{ij}\sim\frac{\phi\epsilon}{c^2\rho^{2/3}}
\left(\frac{p^i p^j}{p^2}-\frac{\delta^{ij}}{3}\right),
\end{equation*}
where $\mathbf{p}$ and $\epsilon$ are the roton momentum and energy.
Total quadrupole moment density is therefore
\begin{equation}
\label{qroton}
{\cal Q}^{ij}
=\int
q^{ij}
n(\epsilon-\mathbf{p}\wb)
\frac{\D\mathbf{p}}{(2\pi\hbar)^3}
\sim
%
%\frac{\phi}{c^2\rho^{2/3}}
%\int
%n(\epsilon-\mathbf{p}\wb)
%\left(\frac{p^i p^j}{p_0^2}-\frac{\delta^{ij}}{3}\right)
%\frac{\epsilon\D\mathbf{p}}{(2\pi\hbar)^3}=
%
\frac{\phi\rn\Delta}{T\rho^{2/3}}
\left(\frac{w^iw^j}{5c^2} - \frac{\delta^{ij}w^2}{15c^2}\right),
\end{equation}
where $\epsilon\approx\Delta$ is the roton energy gap.
Here we assumed Boltzmann distribution $n(\epsilon)$ for rotons and
by \rn\ imply purely roton contribution to the normal density\mycite{LL9}
$\rn=
\int
n(\epsilon)
p^2
\D\mathbf{p}\left/\left(24\pi^3\hbar^3T\right)\right.$.
Equation \eqref{qroton} implies nontrivial
orientation dependence of the contact potential.

%%%%%%%%%%%%%%%%%%%%%%%%%%%%%%%%%%%%%%%%%%%%%%%%%%%%%%%%%%%%%%%%%%%%%%%%%%%

\section{Discussion}
\label{discuss}
Recent experiments\mycite{ryba} demonstrate that a second-sound wave
gives rise to an electric field in superfluid. We can estimate the
expected effect in these experiments from the equations \eqref{PofPT}
and \eqref{qroton}.\footnote{Scalar quadrupole contribution (as described by
Eq.\ref{PofRHON}) is unobservable with external probes: contact potential
cancels the field outside of the liquid.}
Neither pressure
nor total density gradients are present in a second sound wave and the
relative velocity amplitude $w'$ can be obtained from the temperature
amplitude $T'$ as $w'=ST'\left/\left(c_2\rn\right)\right.$, where $c_2$ is the second sound velocity.

The gravitoelectric contribution to the voltage across a second sound wavelength is then
\begin{equation}
U_g =- \int \mathbf{E} \D \mathbf{x}
%= \int 4\pi P \D x
\sim \int 4\pi P_g \D x
\sim\\ - 4\pi
   \frac{\gamma}{3}C T'
\sim
-   \frac{(\varepsilon - 1) M}{12e}
C T',
\end{equation}
while the contact potential contribution is given by
\begin{equation}
U_c \sim
\frac{\pi}{180}
\frac{Ze\rho^{1/3}}{M^{1/3}}
\frac{S^2T'^2}{\rho \rn c^2 c_2^2}\frac{\Delta}{T}.
\end{equation}
The later is quadratic with respect to the relative velocity (and therefore to
the temperature amplitude) and oscillates at the double frequency.

%Complicated experimental cell geometry does not allow to perform direct
%verification of obtained theoretical results, but order of magnitude
%comparison gives satisfactory correspondence.  The ratio $T'/\Delta U$
%measured in the experiment does not depend on the temperature, this fact
%is not explained by the present theory and requires further
%investigation. Failure to observe the effect in the first-sound
%experiment could be due to the fact that walls of the cell are the
%acceleration nodes for the first-sound standing wave.

\section*{Acknowledgements}

I am grateful to A.F.\,Andreev, K.O.\,Keshishev, V.I.\,Marchenko, and
G.E.\,Volovik for fruitful discussions. I would also like to thank
University of New Mexico for its hospitality. This work was supported in
parts by RFBR grants 06-02-17369, 06-02-17281 and RF president program 7018.2006.2.

\end{document}